\documentclass{article}
\usepackage{ijcai26}

\usepackage{times}
\usepackage{soul}
\usepackage{url}
\usepackage[hidelinks]{hyperref}
\usepackage[utf8]{inputenc}
\usepackage[small]{caption}
\usepackage{graphicx}
\usepackage{amsmath}
\usepackage{amsthm}
\usepackage{booktabs}
\usepackage{algorithm}
\usepackage[switch]{lineno}
\usepackage{amsfonts}
\usepackage{algpseudocode}

\title{MARS-DA: A Hierarchical Reinforcement Learning Framework for Risk-Aware Multi-Agent Bidding in Power Grids}

\author{
    Jiayi Chen, Xuan Zhang, Guiling Wang
    \affiliations
    Department of Computer Science, New Jersey Institute of Technology
    \emails
    \{jc2693, xz296, guiling.wang\}@njit.edu
}

\begin{document}

\maketitle

\begin{abstract}
The increasing penetration of renewable energy has introduced substantial volatility into wholesale electricity markets, complicating the optimal bidding strategies for power producers. Traditional Reinforcement Learning (RL) approaches often struggle to balance profit maximization with risk management, frequently overfitting to specific market conditions or failing to account for the stochastic spread between Day-Ahead (DA) and Real-Time (RT) settlements. To address these challenges, this paper makes two primary contributions. First, we introduce and open-source a high-fidelity gymnasium environment for two-settlement electricity market bidding. Grounded in extensive empirical data from the PJM Interconnection, the environment explicitly models the interplay between DA commitments and RT deviations, providing a standardized testbed for general and risk-sensitive agents. Second, we propose \textbf{MARS-DA} (Multi-Agent Regime-Switching for Day-Ahead markets), a novel hierarchical framework that orchestrates distinct sub-policies for risk management and profit seeking. MARS-DA utilizes a top-level Meta-Controller to dynamically blend the actions of two specialized base agents: a ``Safe Agent'' that optimizes for reliable DA allocation and a ``Speculator Agent'' that targets volatile RT arbitrage opportunities. Extensive experiments demonstrate that MARS-DA achieves superior risk-adjusted returns compared to state-of-the-art baselines while maintaining robust regime alignment during periods of extreme market volatility.
\end{abstract}


\section{Introduction}
\label{sec:intro}

Modern electricity bidding is a complex sequential decision problem driven by price volatility and renewable intermittency. Two-settlement designs couple Day-Ahead (DA) commitments with Real-Time (RT) outcomes, exposing participants to significant intertemporal risk. While classical model-based approaches using stochastic or robust optimization are effective under well-defined models, they lack scalability and adaptability in dynamic strategic environments. Conversely, recent advances in reinforcement learning (RL) offer data-driven alternatives, yet most current methods use flat structures and fixed policies that are prone to overfitting and struggle with regime shifts.

A critical challenge remains: bidding requires balancing conservative DA commitments for risk control with opportunistic RT actions to exploit volatility. This tension necessitates hierarchical, risk-aware structures that can adapt to evolving market regimes. 

To address this challenge, we propose \textbf{MARS-DA} (\underline{M}ulti-\underline{A}gent \underline{R}egime-\underline{S}witching for \underline{D}ay-\underline{A}head markets), a hierarchical and risk-aware multi-agent reinforcement learning framework for strategic bidding in two-settlement electricity markets. 
MARS-DA employs a top-level Meta-Controller to dynamically blend the actions of specialized sub-policies, enabling adaptive coordination between profit-seeking and risk management behaviors. 
In addition, we introduce and open-source a high-fidelity Gymnasium-compatible environment grounded in empirical data from PJM Interconnection, explicitly modeling the coupling between day-ahead commitments and real-time deviations. Specifically, this paper makes the following primary contributions:
\begin{enumerate}

    \item We develop an open-source, data-driven simulation environment that captures the two-settlement electricity market structure and explicitly models the interaction between day-ahead and real-time bidding decisions.
    \item We propose a hierarchical multi-agent reinforcement learning framework with adaptive risk-aware policy blending for strategic bidding under market uncertainty.
    \item Extensive experiments demonstrate that MARS-DA achieves superior risk-adjusted performance and improved stability compared to state-of-the-art baselines, particularly during periods of extreme market volatility.
\end{enumerate}

\section{Related Work}
\label{sec:related_work}

Optimal bidding in electricity markets has been studied from optimization and learning perspectives. We review four related areas: (i) model-based bidding under uncertainty, (ii) deep reinforcement learning for market bidding, (iii) hierarchical and risk-aware reinforcement learning, and (iv) simulation environments.

\subsection{Evolution of Bidding Strategies}

Electricity markets typically operate under a two-settlement structure with a Day-Ahead (DA) market followed by Real-Time (RT) balancing, creating a sequential decision problem under price and system uncertainty. Classical model-based approaches formulate bidding as optimization under uncertainty, including stochastic programming~\cite{heredia2012stochastic,herding2023stochastic}, robust optimization~\cite{sun2017robust,nemati2024flexible}, and rolling-horizon methods~\cite{feng2023multi,chen2024rolling}, with extensions to coordinated participation across energy and reserve markets. Despite their effectiveness, these methods rely on explicit uncertainty models and repeated re-optimization, limiting their adaptability to dynamic market environments.

Recent advances in deep reinforcement learning (DRL) provide a data-driven alternative, enabling agents to learn bidding strategies directly from market interactions. DRL has been applied to DA bidding and joint energy-regulation market participation~\cite{di2025reinforcement,anwar2022proximal}, while multi-agent RL (MARL) addresses strategic interactions via Nash equilibrium approximation and knowledge transfer~\cite{du2021approximating,wu2022strategic}. However, most existing DRL methods learn fixed policies that struggle under regime shifts and offer limited control over risk-return tradeoffs.

\subsection{Hierarchical and Risk-Aware RL}
Hierarchical reinforcement learning has recently been explored for electricity market applications to address multi-stage decision-making and long planning horizons. Zhang et al.~\cite{zhang2025arbitrage} show that hierarchical multi-agent reinforcement learning can improve profitability and stability in local electricity market arbitrage by separating high-level coordination from low-level trading actions. In the financial domain, Chen et al.~\cite{chen2025mars} propose MARS, a meta-adaptive framework that orchestrates heterogeneous risk-specialized agents through a high-level controller for portfolio management, demonstrating superior drawdown control during market stress.
Risk-sensitive reinforcement learning incorporates risk measures (e.g., CVaR) to address tail risks. Distributional RL and CVaR-based objectives have been shown to yield adjustable risk-averse policies, though typically with fixed risk measures that limit adaptability across regimes \cite{heche2025risk,lim2022distributional,stanko2019risk}. 

\subsection{Simulation Environments for Power Systems}
Standardized environments are important for reproducible evaluation of learning-based methods in power systems. Grid2Op provides a realistic sequential decision-making testbed for grid operations and has been widely adopted \cite{donnot2020grid2op}. RL2Grid builds on Grid2Op to offer standardized benchmark tasks facilitating fair comparisons across RL algorithms \cite{marchesini2025rl2grid}. 
However, existing environments mainly focus on grid operations or single-market participation and typically do not model coordinated DA--RT bidding under a two-settlement market structure, which motivates the market environment design in our work.

\begin{figure*}[h]
    \centering
\includegraphics[width=0.8\linewidth]{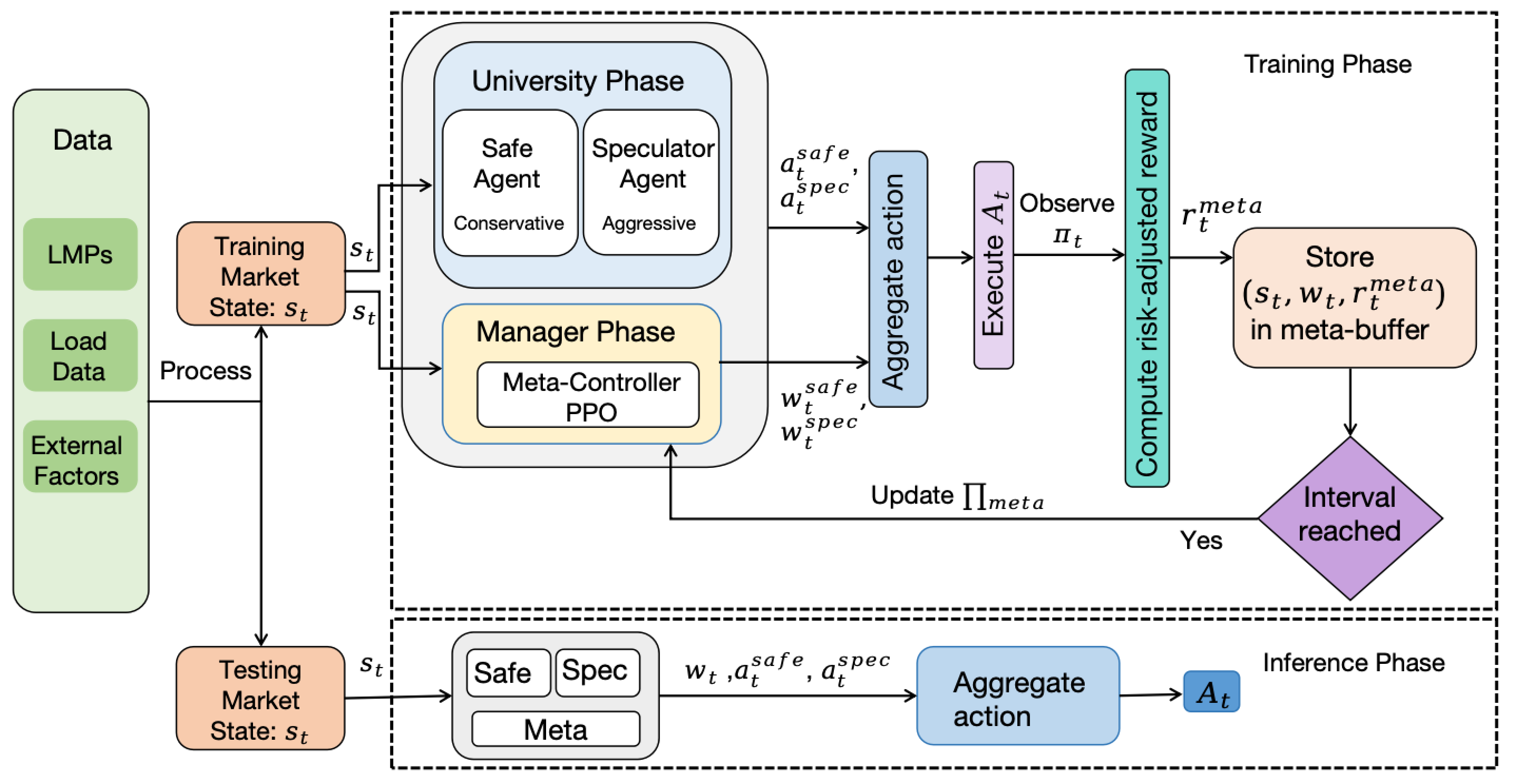}
    \caption{The MARS-DA Hierarchical Framework. 1) During training, the Meta-Controller (Manager) observes the market state $s_t$ to dynamically weight the proposals of the pre-trained Safe and Speculator agents (Workers). 2) The aggregated action is executed in the two-settlement electricity market environment, where risk-adjusted rewards are computed and used to update $\Pi_{meta}$. 3) During inference, the learned policy adaptively blends sub-agent actions to produce the final day-ahead allocation decision.}
    \label{fig:framework}
\end{figure*}

\section{Problem Formulation}
\label{sec:problem}

We consider a strategic generator participating in a two-settlement wholesale electricity market comprising a Day-Ahead (DA) market and a Real-Time (RT) market. The generator's objective is to maximize cumulative risk-adjusted profit by optimizing its capacity allocation between these two markets.

\subsection{Market Mechanics}

The market-clearing process follows a standard two-stage settlement mechanism. Any capacity not committed to the DA market ($q_{DA}$) is automatically made 
available to the Real-Time market ($q_{RT} = P_{max} - q_{DA}$), representing 
the speculative portion of the bid.


\subsubsection{Day-Ahead (DA) Market}

In the DA market, which clears 24 hours prior to the operating day, the generator submits a quantity offer $q_{DA}$ for each hour $t$. The Independent System Operator (ISO) clears the market, resulting in a Day-Ahead Locational Marginal Price ($\lambda_{t}^{DA}$). The generator is financially binding to this schedule and receives revenue $\lambda_{t}^{DA} \cdot q_{DA}$.

\subsubsection{Real-Time (RT) Market}
The RT market balances deviations between the DA schedule and actual real-time generation. Any capacity not committed to the DA market ($q_{DA}$) is automatically made available to the Real-Time market ($q_{RT} = P_{max} - q_{DA}$), representing the speculative portion of the bid. This quantity is settled at the Real-Time Locational Marginal Price (LMP), denoted as $\lambda_{t}^{RT}$.

The total financial profit $\pi_{t}$ for hour $t$ accounts for revenues across both settlements, variable marginal costs, unit commitment costs, and operational penalties:



\begin{equation}
\begin{aligned}
\pi_{t} = & (\lambda_{t}^{DA}q_{DA} + \lambda_{t}^{RT}q_{RT}) - C_{marg}(q_{DA} + q_{RT}) \\
& - C_{startup} \cdot \mathbb{I}_{start} - \Omega_{penalty}
\end{aligned}
\end{equation}


where $C_{marg}$ is the marginal fuel cost per unit of generation, $C_{startup}$ represents the cost incurred when the generator transitions from an offline to an online state with $\mathbb{I}_{start}=1$, and $\Omega_{penalty}$ represents financial fines for violating physical constraints, such as ramping limits or minimum up/down time requirements. 


This formulation captures the strategic trade-off between locked-in DA revenue and volatile RT arbitrage, while accounting for the inter-temporal costs and physical risks of power generation.

\subsection{Markov Decision Process (MDP)}

We model the strategic bidding problem as a finite-horizon Markov Decision Process (MDP) defined by the tuple $(\mathcal{S}, \mathcal{A}, P, R, \gamma)$.




\subsubsection{State Space ($\mathcal{S}$)}
The state $s_t \in \mathcal{S}$ captures market conditions and the physical state of the generation asset, defined as $s_t = [M_t, U_t, \tau_t]$. The \textbf{Market State} ($M_t$) includes the current Day-Ahead price ($P_{DA}$), the 24-hour rolling volatility signal ($\sigma_{24h}$), and the load forecast ($L_t$). The \textbf{Unit State} ($U_t$) manages physical constraints via the binary commitment status ($u_t \in \{0,1\}$), duration counters for Minimum Up/Down time compliance, and the previous power output level. Finally, \textbf{Time Features} ($\tau_t$) utilize cyclical sine/cosine encoding of the hour and day to capture seasonality.







\noindent \textbf{Action Space ($\mathcal{A}$):} 

The agent outputs a continuous action $a_{raw} \in [-1, 1]$, which is mapped to the Day-Ahead Allocation Ratio $\alpha_t \in [0, 1]$ via $\alpha_t = (a_{raw} + 1) / 2$. The physical quantities committed to each market are then derived as $q_{DA} = \alpha_t \cdot P_{max}$ and $q_{RT} = (1 - \alpha_t) \cdot P_{max}$. This strategic variable determines the hedge ratio: any capacity not locked in the DA market is automatically split into the RT market, representing the speculative portion.

\subsubsection{Reward Function ($R$)}
The environment calculates the raw financial profit $\pi_{t}$ at each time step $t$ according to Eq. (1), incorporating settlement revenues, marginal fuel costs ($C_{marg}$), startup costs ($C_{startup}$), and penalties ($\Omega_{penalty}$) for constraint violations. While driven by this financial signal, MARS-DA utilizes specialized reward shaping: \textbf{Base Agents} (Safe and Speculator) optimize specialized profit-share rewards to induce role specialization (see Section 4.3), while the \textbf{Meta-Controller} optimizes a general risk-adjusted return $r_{meta} = \pi_{t} - \lambda_{risk} \cdot \text{Risk}(\pi_{t})$. In practice, we instantiate this risk term using the concave utility defined in Eq. (6). Here, $\lambda_{risk}$ is the risk aversion coefficient controlling the trade-off between maximizing profit and minimizing return variance or downside risk. This formulation encourages consistent, reliable gains over volatile, high-risk strategies.
\section{Methodology: The MARS-DA Framework}
\label{sec:methodology}




We propose the Multi-Agent Regime-Switching Day-Ahead (MARS-DA) framework, a hierarchical reinforcement learning system designed to optimize bidding strategies in two-settlement electricity markets. The framework decouples the complexity of market participation into two distinct layers: specialized base agents that propose regime-specific actions, and a Meta-Controller that dynamically blends these proposals to maximize risk-adjusted returns.

\subsection{Overall Training Algorithm}

The complete training and evaluation procedure for the MARS-DA framework is summarized in Algorithm \ref{alg:mars_training}.

\subsection{Hierarchical Architecture}

The system operates as a hierarchical ``Manager-Worker" structure, where all components are trained using Proximal Policy Optimization (PPO). At each time step $t$, the system observes the market state $s_t$:

\begin{enumerate}
    \item \textbf{Workers (Base Agents):} Two independent agents, trained to specialize in distinct market objectives, process $s_t$ and propose allocation actions $a_{t}^{(k)}$.
    \item \textbf{Manager (Meta-Controller):} A higher-level policy network $\Pi_{meta}(s_t)$ outputs a blending weight vector $\mathbf{w}_t = [w_t^{safe}, w_t^{spec}]$ such that $w_t^{safe} + w_t^{spec} = 1$.
    \item \textbf{Execution:} The final action submitted to the environment is the weighted linear combination of the proposals:
    \begin{equation}
        A_t = w_t^{safe} \cdot a_{t}^{safe} + w_t^{spec} \cdot a_{t}^{spec}
    \end{equation}
\end{enumerate}

This architecture allows the system to seamlessly transition between conservative and aggressive postures without retraining the underlying policies.





\subsection{Heterogeneous Base Agents}

The base agents are trained in a pre-training phase (``University Phase") using \textbf{profit-aware role enforcement}. Unlike pure behavior cloning, which only enforces volume allocation, this method ensures that agents learn to be \textit{profitable} within their assigned market role. This prevents the "blind volume" problem, where an agent might secure market share even when it results in financial losses.

\subsubsection{The Safe Agent (Day-Ahead Specialist)}



The Safe Agent is trained to maximize profit specifically generated through the Day-Ahead (DA) market. Its reward function scales the realized financial profit $\pi_{t}$ by the DA allocation ratio $\alpha_{t}$, while heavily penalizing any utilization of the Real-Time (RT) market:

\begin{equation}
    r_{safe} = (\pi_{t} \cdot \alpha_{t}) - (|\pi_{t}| \cdot (1 - \alpha_{t}) \cdot \lambda_{role})
\end{equation}


Functionally, this acts as a profit-seeking DA Machine. It is incentivized to trade only when profitable, but strictly within the DA mechanism. The penalty term ensures that even if the RT market is profitable, the Safe agent is discouraged from accessing it, thereby enforcing role specialization during the University Phase.

\subsubsection{The Speculator Agent (Real-Time Specialist)}


The Speculator Agent is trained to maximize profit specifically derived from Real-Time (RT) exposure. Its reward function scales the realized financial profit $\pi_{t}$ by the RT allocation ratio $(1 - \alpha_{t})$, penalizing any capacity locked in the DA market:


\begin{equation}
    r_{spec} = (\pi_{t} \cdot (1 - \alpha_{t})) - (|\pi_{t}| \cdot \alpha_{t} \cdot \lambda_{role})
\end{equation}


This agent learns aggressive arbitrage strategies, effectively acting as an RT Machine. It is incentivized to withhold capacity from the DA market—not just to withhold, but to capture the higher value anticipated in the RT settlement. By applying the role penalty $\lambda_{role}$, the agent learns to prioritize speculative opportunities while remaining profitable within its assigned market role.

\subsection{Meta-Adaptive Controller}

The Meta-Controller is the strategic core of MARS-DA. It is a PPO-based policy network that learns to blend the frozen base agents to optimize a global financial objective.

\subsubsection{Objective Function}

The Meta-Controller is trained using Proximal Policy Optimization (PPO) to maximize a risk-adjusted return objective. We employ the clipped surrogate objective function to ensure stable policy updates:


\begin{equation}
\resizebox{0.9\linewidth}{!}{%
$L^{CLIP}(\theta) = \mathbb{E}_t [\min(r_t(\theta) \hat{A}_t, \text{clip}(r_t(\theta), 1-\epsilon, 1+\epsilon) \hat{A}_t)]$
}
\end{equation}

where $r_t(\theta)$ is the probability ratio, $\hat{A}_t$ is the generalized advantage estimate (GAE), and $\epsilon=0.2$ is the clipping hyperparameter.

\subsubsection{Risk-Adjusted Reward Shaping}
To stabilize training and induce risk-averse behavior, we implement a Concave Utility Reward rather than a simple linear profit target. The reward function $r_{meta}$ penalizes the squared magnitude of the realized profit, effectively discouraging ``jackpot-seeking" strategies that often carry catastrophic downside risk:


\begin{equation}
    r_{meta} = \frac{\pi_{t}}{S_{linear}} - \frac{\lambda_{risk}}{2} \cdot \left(\frac{\pi_{t}}{S_{var}}\right)^{2}
\end{equation}




where $\pi_t$ is the realized profit at time $t$. We introduce two scaling factors: $S_{linear}=1000$ normalizes the linear profit term, while $S_{var}=100$ scales the profit within the quadratic penalty term. With the risk aversion coefficient $\lambda_{risk}$ set to 5.0, this formulation penalizes the \textit{magnitude} of outcomes. By subtracting a term proportional to $\pi_t^2$, the objective function becomes concave, effectively serving as a proxy for minimizing the variance of returns. This encourages the Meta-Controller to avoid ``jackpot-seeking" behaviors—where high profits are accompanied by high volatility—and instead favor consistent, reliable gains typically provided by the Safe agent.

\begin{algorithm}[t]
\caption{MARS-DA Training Procedure}
\label{alg:mars_training}
\begin{algorithmic}[1]
\small 
\State \textbf{Input:} Training data $\mathcal{D}_{train}$, max steps $T_{base}, T_{meta}$
\State \textbf{Initialize:} Workers $\pi_{safe}, \pi_{spec}$; Manager $\Pi_{meta}$
\Statex \textbf{Phase 1: University Phase (Pre-train Workers)}
\For{$k \in \{safe, spec\}$}
    \State Train $\pi_k$ on $s_t \in \mathcal{D}_{train}$ using specialized $r_t^k$ (Eqs. 3-4) until $T_{base}$. 
    \State Freeze parameters of $\pi_k$. 
\EndFor
\Statex \textbf{Phase 2: Manager Phase (Train Meta-Controller)}
\While{steps $< T_{meta}$}
    \State Get base proposals $a_t^{safe}, a_t^{spec}$ and mixing weights $\mathbf{w}_t \sim \Pi_{meta}(s_t)$. 
    \State Execute $A_t = \sum w_t^k a_t^k$; observe $\pi_t$ and compute $r_t^{meta}$ (Eq. 6). 
    \State Store $(s_t, \mathbf{w}_t, r_t^{meta})$; update $\Pi_{meta}$ via PPO Clipped Objective. 
\EndWhile
\State \textbf{Output:} Trained $\Pi_{meta}$ and frozen base agents $\pi_k$. 
\end{algorithmic}
\end{algorithm}

\section{The MARS-DA Environment}
\label{sec:environment}

A critical contribution of this work is the development of \texttt{StrategicBiddingEnv}, an open-source, Gymnasium-compatible reinforcement learning environment designed to simulate the dynamics of two-settlement electricity markets with high fidelity. Unlike existing simplified benchmarks, this environment integrates real-world historical data with a realistic physical generator model, enabling the training of agents that can generalize to complex market regimes.

\subsection{Data Pipeline and Reconstruction}

The environment is built upon a comprehensive dataset of historical market conditions from the PJM Interconnection, spanning a five-year period from 2018 to 2022, and is further augmented with a recent testing period from September 2024 to September 2025. The data pipeline aggregates multiple independent sources to construct a rich state space for the agent.


\subsubsection{Data Sources}
We utilize the PJM DataMiner API to ingest hourly time-series data, including Day-Ahead (DA) and Real-Time (RT) Locational Marginal Prices (LMPs) to model price spreads and volatility. To provide demand-side context, we include metered actual load and day-ahead load forecasts. This dataset is further augmented with external weather data (temperature and wind speed) from Open-Meteo and daily natural gas spot prices from the EIA, which serve as leading indicators of electricity price fluctuations.


\subsubsection{Data Processing}

Raw data was reconstructed using linear interpolation for short gaps ($<$4 hours) and seasonal averaging for longer intervals. To prevent leakage, we employed a strict chronological split: 2018--2021 for training, 2022 for Test Period 1, and September 2024--September 2025 for Test Period 2.

\subsection{Simulation Logic}

The simulation logic models the financial and physical clearing of a single gas-fired generator participating in the market.

\subsubsection{Generator Model}
The agent controls a hypothetical unit defined by physical parameters: capacity ($P_{max} = 100$ MW), minimum stable generation ($P_{min}$), and ramp rates ($RR$). The unit has a marginal cost $MC$ derived from the daily gas price, simulating the variable operating cost.

\subsubsection{Market Clearing Process}

At each step $t$, the environment emits a state vector $s_t$ containing 24-hour historical LMPs, load forecasts, and unit status. The agent submits an allocation $a_t \in [0, 1]$, setting the Day-Ahead quantity $q_{DA} = a_t P_{max}$. Upon retrieving realized DA and RT prices ($\lambda_{t}^{DA}, \lambda_{t}^{RT}$), the system calculates the financial outcome: $q_{DA}$ is settled at $\lambda_{t}^{DA}$, while the remainder ($q_{RT} = P_{max} - q_{DA}$) is settled at $\lambda_{t}^{RT}$. Finally, the time index advances ($t \leftarrow t+1$) and the state buffer rolls forward.

This simulation explicitly models the \textit{opportunity cost} of locking in DA prices versus exposing capacity to RT volatility, capturing the core risk-return trade-off of the merchant generation business.

\section{Experimental Setup}
\label{sec:setup}

We evaluate the MARS-DA framework using a rigorous benchmarking suite designed to test performance consistency, risk management, and adaptability across distinct market regimes. Our experimental design includes a comprehensive ablation study to isolate the contribution of each architectural component.




\subsection{Dataset and Splitting}
We utilize PJM Interconnection data (2018--2025) across three periods. \textbf{Training (2018--2021, $N \approx 35,000$)} serves as the University Phase for training all agents. \textbf{Test Period 1 (2022, $N = 8,760$)} provides a hold-out set characterized by high fuel prices and post-pandemic recovery. \textbf{Test Period 2 (Sep 2024--Sep 2025, $N \approx 9,500$)} evaluates adaptation to recent structural shifts, such as increased renewable penetration, which were absent from the training distribution.

\subsection{Baselines}

We benchmark MARS-DA against four representative baselines:

\begin{itemize}
    \item \textbf{NNSF (Vanilla PPO):} A monolithic agent trained end-to-end. Represents standard Deep RL without hierarchy.
    \item \textbf{Vanilla SAC:} An off-policy Soft Actor-Critic agent maximizing entropy. Benchmarks sample efficiency.
    \item \textbf{CVaR-PPO (Risk-Averse):} A PPO variant optimizing Conditional Value at Risk ($\alpha=0.05$). Benchmarks risk-sensitive RL.
    \item \textbf{RollingOpt (MPC Heuristic):} A Model Predictive Control baseline using 24-hour moving average forecasts. Represents non-learning industry standards.
\end{itemize}

\subsection{Evaluation Metrics}
Results are averaged over 10 independent runs (200 episodes each) per test period. We evaluate performance using the \textbf{Sharpe Ratio} for risk-adjusted returns, \textbf{Sortino Ratio} for downside-specific risk, \textbf{Maximum Drawdown (MDD)} to measure peak-to-trough portfolio declines, and a \textbf{Regime Alignment Score} to measure the correlation between the meta-controller's weights and the optimal ex-post strategy.


\section{Results and Analysis}
\label{sec:results}

\subsection{Comparative Performance}
Table \ref{tab:performance_stacked}
presents performance metrics for MARS-DA compared with baseline strategies across two distinct test periods.
In Test Period 1 (2022), MARS-DA demonstrates superior risk-adjusted performance, achieving the highest Sharpe Ratio (0.978) and the lowest Maximum Drawdown (2.84\%). This result validates the Meta-Controller's ability to preserve capital during market stress while capturing upside during stable periods.
In Test Period 2 (2024-2025), despite the significantly higher market volatility, MARS-DA outperformed all baselines. It achieved the highest cumulative return (\$4.45M) and Sharpe Ratio (0.270) while maintaining a remarkably low maximum drawdown of 8.12\%, compared to $>$55\% for the baseline agents. This result highlights the Meta-Controller's critical role in mitigating ``tail risks" that are catastrophic for single-policy agents during regime shifts.

\begin{figure}[t]
    \centering
    \includegraphics[width=\columnwidth]{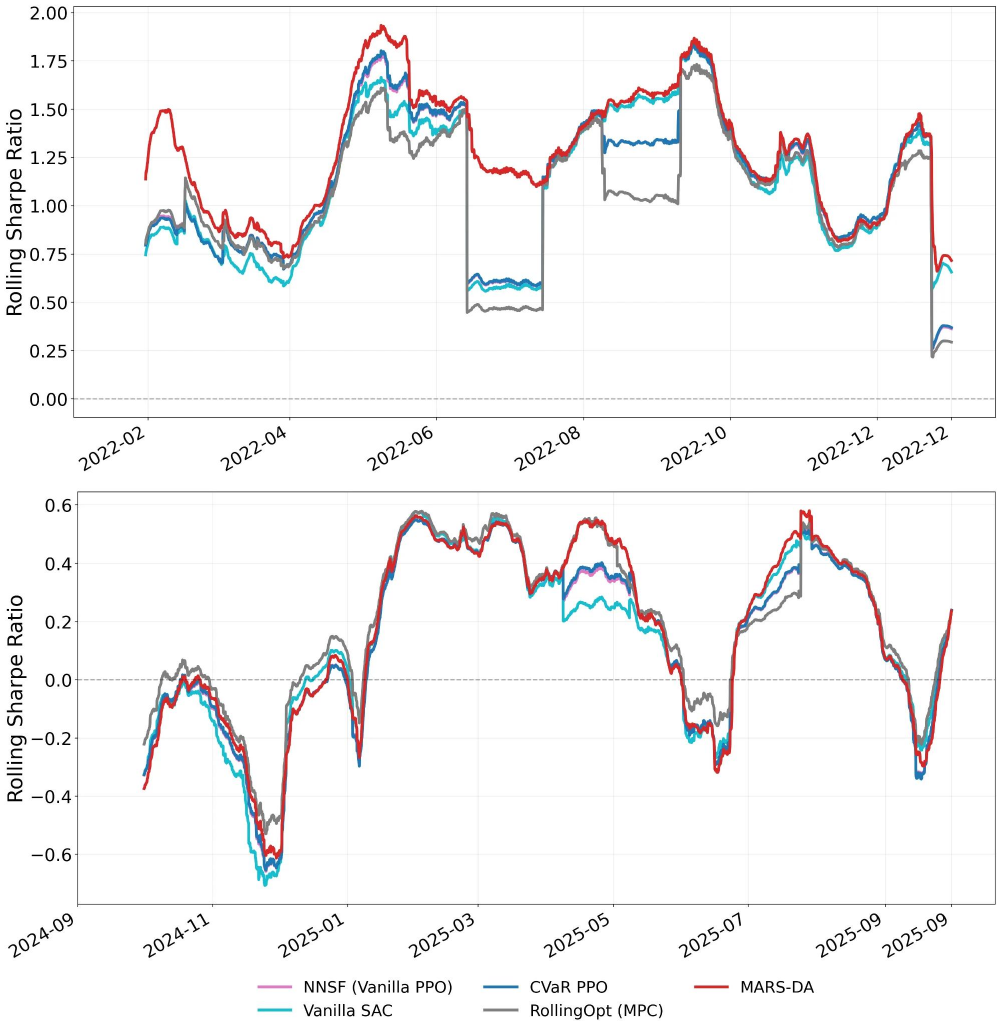}
    \caption{Rolling Sharpe Ratio comparison. Top: Test Period 1 (2022). Bottom: Test Period 2 (2024-2025). MARS-DA (red) maintains consistently higher risk-adjusted returns.}
    \label{fig:rolling_sharpe}
\end{figure}

\begin{figure}[t]
    \centering
    \includegraphics[width=\columnwidth]{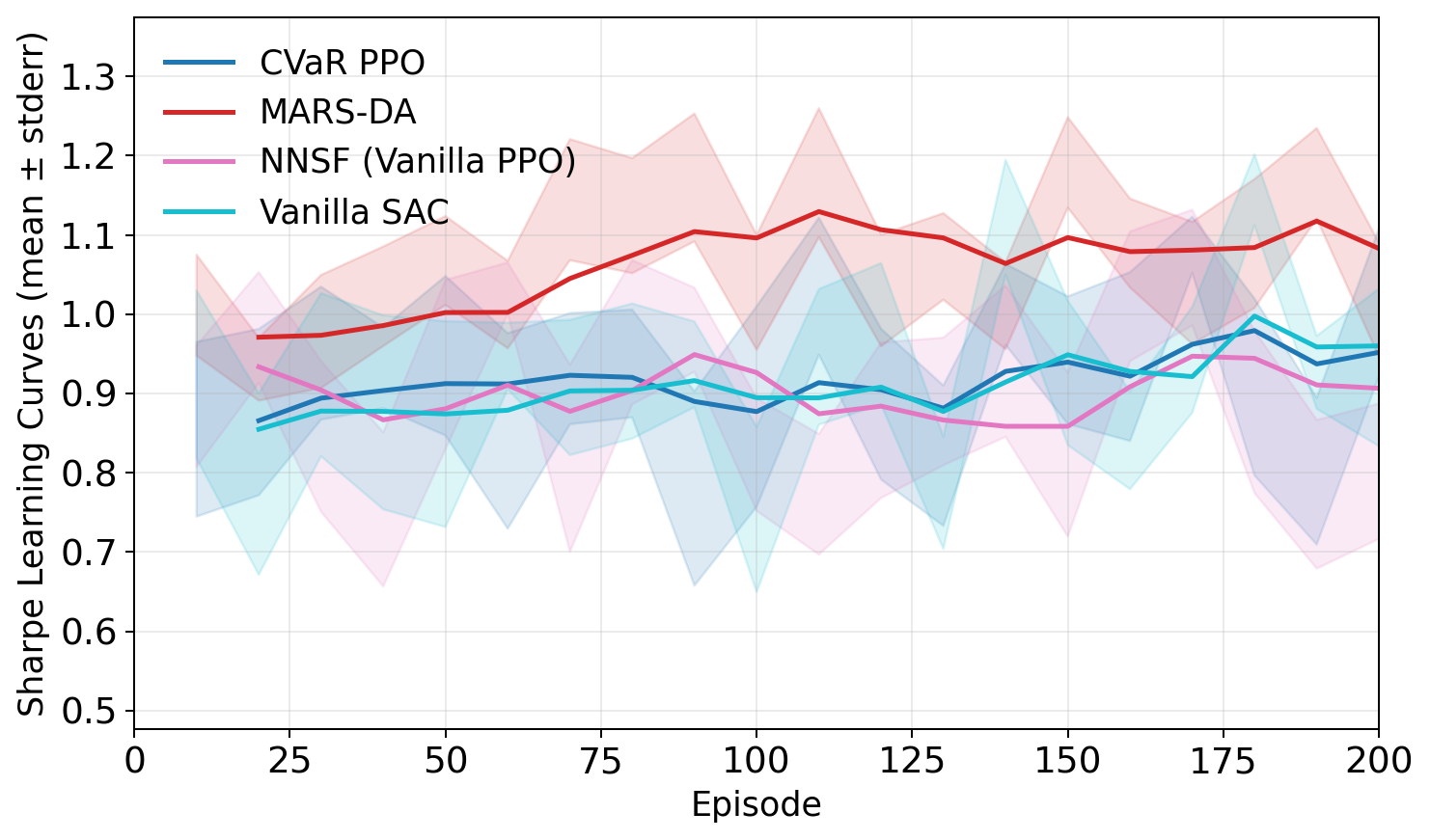}
    \caption{Sharpe Ratio learning curves (mean $\pm$ stderr over 10 seeds). MARS-DA achieves faster convergence and higher asymptotic performance.}
    \label{fig:learning_curve}
\end{figure}


\begin{table}[t]
\centering
\caption{Performance comparison across Test Period 1 (2022) and Test Period 2 (2024–2025).}
\label{tab:performance_stacked}
\resizebox{\columnwidth}{!}{%
\begin{tabular}{lcccc}
\toprule
\textbf{Method} & \textbf{Cum. Ret.(\$)} & \textbf{Sharpe} & \textbf{Sortino} & \textbf{Max DD (\%)} \\
\midrule
\multicolumn{5}{l}{\textit{Panel A: Test Period 1 (2022)}} \\
\midrule
\textbf{MARS-DA} & \$60.1M & \textbf{0.978} & 3.142 & \textbf{2.84\%} \\
CVaR PPO & \$61.5M & 0.697 & 3.126 & 3.99\% \\
RollingOpt & \$60.7M & 0.646 & 3.520 & 36.78\% \\
Vanilla SAC & \$62.4M & 0.424 & 3.226 & 3.15\% \\
NNSF (PPO) & \textbf{\$67.6M} & 0.545 & \textbf{3.590} & 6.77\% \\
\midrule
\multicolumn{5}{l}{\textit{Panel B: Test Period 2 (2024–2025)}} \\
\midrule
\textbf{MARS-DA} & \textbf{\$4.45M} & \textbf{0.270} & \textbf{1.808} & \textbf{8.12\%} \\
CVaR PPO & \$2.86M & 0.248 & 1.150 & 55.91\% \\
RollingOpt & \$1.15M & 0.145 & 0.478 & 61.63\% \\
Vanilla SAC & \$1.70M & 0.126 & 0.713 & 55.48\% \\
NNSF (PPO) & \$1.46M & 0.183 & 0.601 & 60.18\% \\
\bottomrule
\end{tabular}%
}
\end{table}

\subsection{Rolling Performance and Learning Dynamics}

Figure~\ref{fig:rolling_sharpe} presents rolling Sharpe Ratio trajectories across both test periods. In 2022 (top), MARS-DA consistently maintains the highest values, while RollingOpt exhibits erratic step-like behavior during mid-year volatility. In 2024-2025 (bottom), all methods face challenging out-of-distribution conditions, with Sharpe ratios frequently dropping below zero, yet MARS-DA achieves the fastest recovery and the highest peaks ($\approx$0.55).

Figure~\ref{fig:learning_curve} reveals the source of this advantage: MARS-DA exhibits rapid convergence and maintains higher Sharpe ($\approx$1.0--1.1) throughout training compared to baselines ($\approx$0.85--0.95), demonstrating that the hierarchical architecture enables efficient learning without sacrificing stability.

\subsection{Regime Alignment Analysis}
\begin{figure}[t]
\centering
\includegraphics[width=\columnwidth]{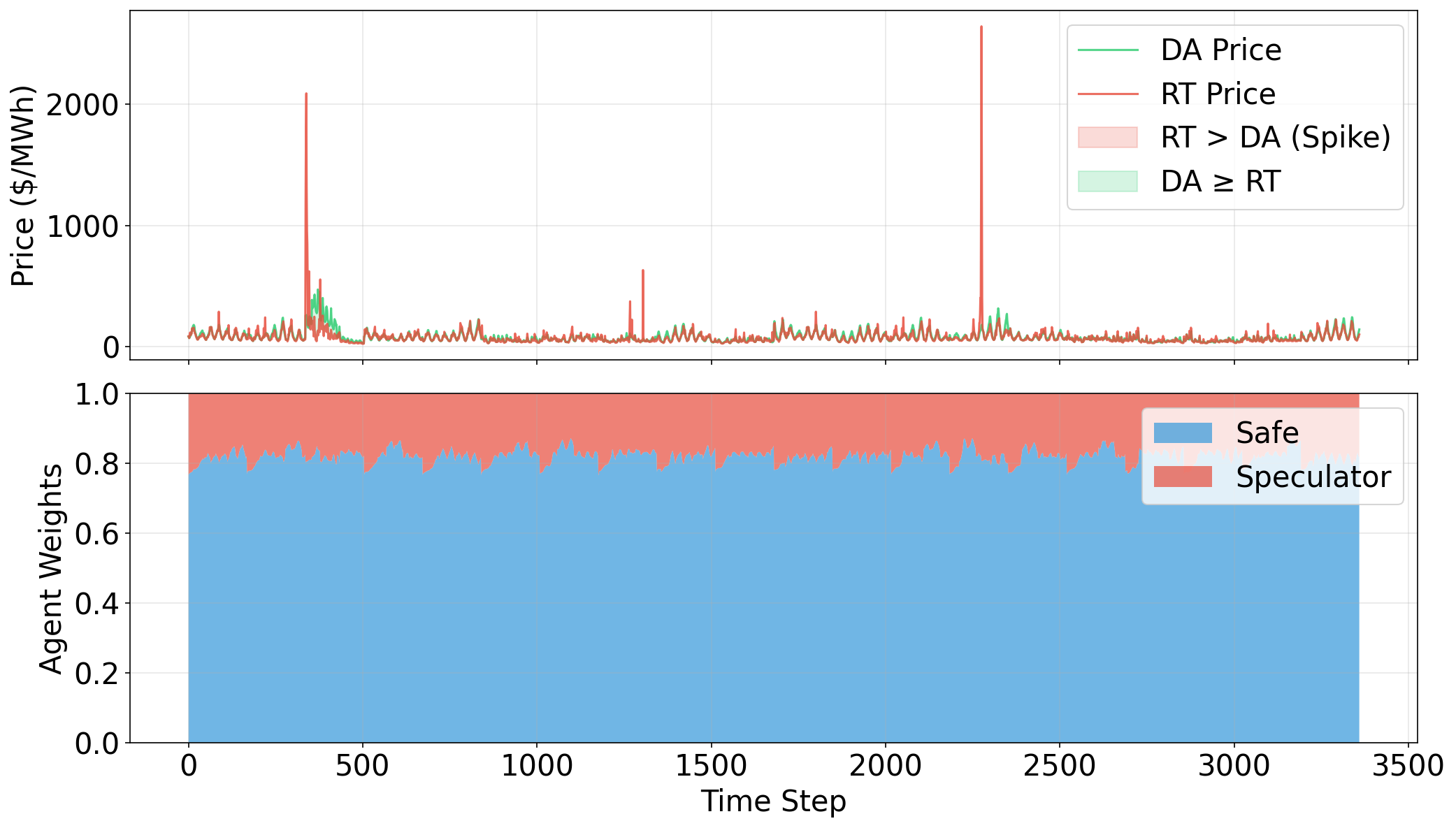}
\caption{Regime Alignment Dynamics. The Meta-Controller (stacked areas) dynamically reallocates weights between Safe (Blue) and Speculator (Orange) agents in response to Market Price Volatility (Red Line), acting as an automated risk-manager.}
\label{fig:regime_switching}
\end{figure}

To understand the source of MARS-DA's superior risk-adjusted returns, we analyzed the Meta-Controller's policy dynamics using two key metrics: Allocation Entropy (measuring the diversity of agent usage) and Regime Alignment Score (measuring the correlation between the Speculator agent's weight and market volatility).
In Test Period 1, MARS-DA exhibited a high Allocation Entropy of 0.517, confirming that the system actively switched between the \textit{Safe} and \textit{Speculator} agents rather than collapsing to a single static policy.
Crucially, the Regime Alignment Score was -0.075. While a positive score would indicate a "momentum" strategy (increasing risk during high volatility), this slightly negative correlation suggests sophisticated contrarian hedging behavior. The Meta-Controller learned to dampen the Speculator's influence during extreme volatility spikes—periods often associated with unpredictable price crashes (negative skewness)—thereby avoiding the catastrophic drawdowns suffered by the standalone Speculator-like baselines. Instead, it deployed the Speculator primarily during moderate volatility regimes where price dispersion provided profitable arbitrage opportunities without excessive tail risk.
Figure \ref{fig:regime_switching} visualizes this behavior, showing the Meta-Controller rapidly shifting weights to the Safe agent (blue area) exactly as price volatility (red line) breaches critical thresholds, effectively acting as an automated ``circuit breaker."

\subsection{Ablation Studies}
To validate the architectural decisions of MARS-DA, we conducted the following ablation experiments (results summarized in Table \ref{tab:ablation}).
\begin{enumerate}
\item \textbf{Heterogeneity vs. Homogeneity:}
We compared the proposed heterogeneous ensemble (Safe + Speculator) against a homogeneous baseline of generic agents (represented here by \textit{Vanilla SAC}). The heterogeneous MARS-DA system achieved a Sharpe Ratio of \textbf{0.978} compared to \textbf{0.424} for the homogeneous baseline. This confirms that performance gains stem from functional role specialization—allowing distinct policies to capture different market regimes—rather than simple variance reduction from ensembling.

\begin{table}[t]
\centering
\caption{Ablation Analysis. The Heterogeneous, Dynamic, Hierarchical system (MARS-DA) outperforms simplifications.}
\label{tab:ablation}
\begin{tabular}{lcc}
\toprule
\textbf{Ablation Configuration} & \textbf{Sharpe} & \textbf{Max Drawdown} \\
\midrule
\textbf{MARS-DA (Full System)} & \textbf{0.978} & \textbf{2.84\%} \\
Homogeneous (Vanilla SAC) & 0.424 & 3.15\% \\
Static Weights (50/50) & 0.458 & 3.69\%  \\
Ensemble Size 3 & 0.601 & 4.04\%  \\
Best-Single (CVaR PPO) & 0.697 & 3.99\% \\
\bottomrule
\end{tabular}
\end{table}

\item \textbf{Dynamic vs. Static Weighting:}
We replaced the learned Meta-Controller with fixed equal weights ($w_{safe}=0.5, w_{spec}=0.5$), forming a static Safe--Speculator ensemble. While this static ensemble achieved higher raw profit (\$77.5M vs.\ \$64.3M), its risk-adjusted performance was much worse (Sharpe $0.458$ vs.\ $0.920$) and it suffered a larger maximum drawdown ($3.69\%$ vs.\ $0.90\%$). This shows that \textit{MARS-DA}'s learned, state-dependent regime switching is crucial for \emph{doubling risk-adjusted returns and sharply reducing downside risk}, rather than merely averaging expert policies.
\item \textbf{Ensemble Size:}
We extended the base architecture to include a third ``Neutral'' agent, forming a 3-agent ensemble. The Neutral agent was designed to maintain balanced DA/RT allocation, with a reward function that penalizes deviations exceeding 20\% from a 50/50 split. As shown in Table~\ref{tab:ablation}, the 3-agent configuration achieved a Sharpe Ratio of 0.601 and a Maximum Drawdown of 4.04\%, compared to 0.978 and 2.84\% for the 2-agent system. The additional agent introduced optimization complexity, hindering convergence, as the Meta-Controller struggled to effectively coordinate three competing strategies. These results confirm that the 2-agent (Safe + Speculator) configuration provides the optimal balance between strategic diversity and training stability.
\item \textbf{Meta-Controller Necessity:}
We compared the full system against the single best-performing base agent (\textit{CVaR PPO}, the ``Best-Single" baseline). While the Best-Single agent was robust (Sharpe 0.697), the full hierarchical system outperformed it (Sharpe 0.978). The Meta-Controller added value by accessing the ``Speculator" agent's aggressive yield during stable periods—upside that the conservative Best-Single agent systematically missed.
\end{enumerate}

\begin{figure}[t]
    \centering
    \includegraphics[width=\columnwidth]{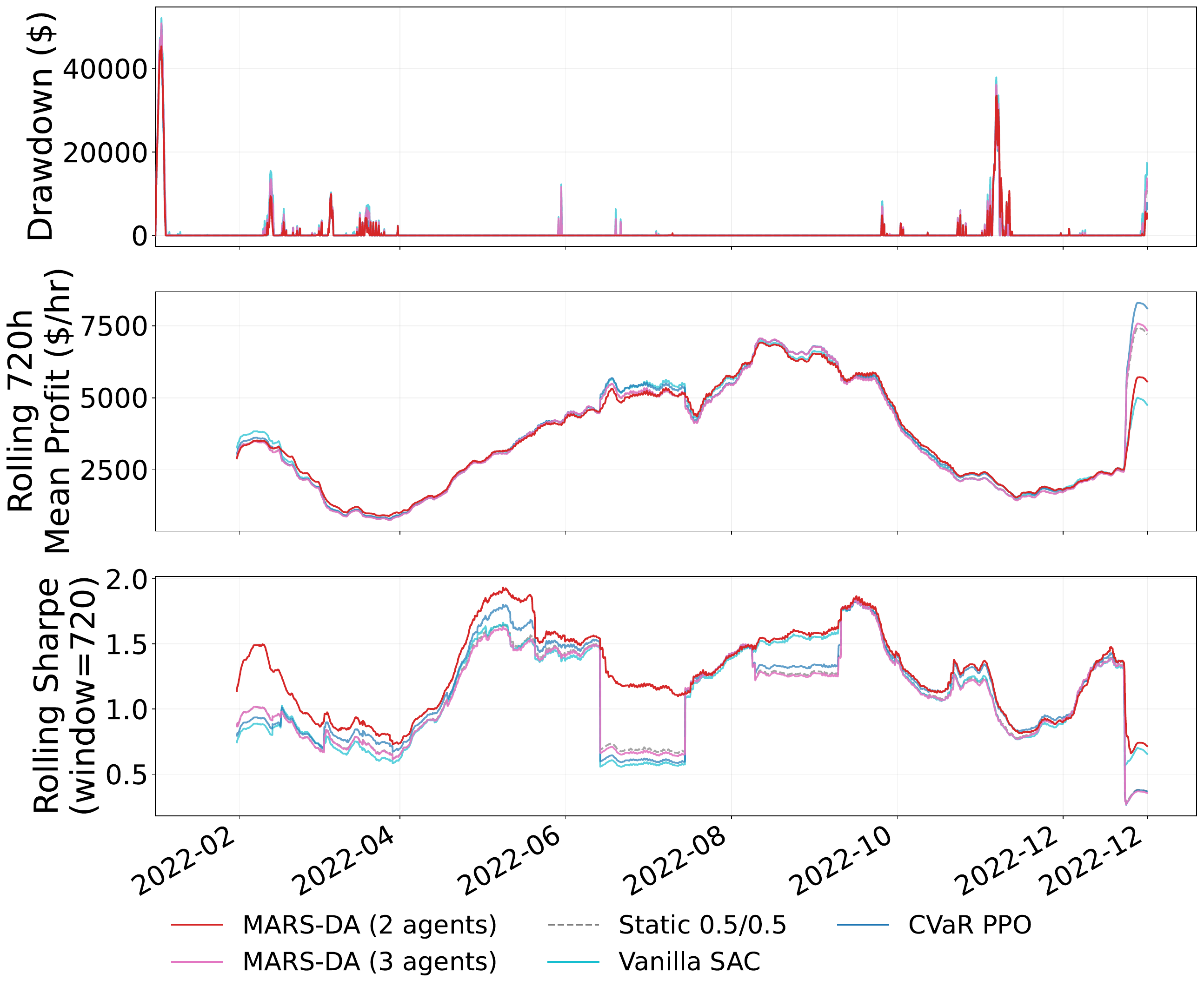}
    \caption{Ablation study visualization on Test Period 1 (2022). From top to bottom: Cumulative Return, Drawdown, Rolling 720h Mean Profit, and Rolling Sharpe Ratio. MARS-DA (2 agents) achieves optimal risk-return balance with consistently lower drawdowns and higher Sharpe ratios.}
    \label{fig:ablation}
\end{figure}

Figure~\ref{fig:ablation} visualizes these ablation variants across four complementary metrics on Test Period 1. The Drawdown panel reveals that MARS-DA (2 agents) maintains consistently low drawdowns throughout the year, while competing methods, particularly the 3-agent variant and CVaR PPO, suffer periodic spikes exceeding \$30K--50K during volatile periods. The Rolling Mean Profit (720h window) shows all methods track similar profit trajectories, yet the Rolling Sharpe panel demonstrates MARS-DA's superior risk-adjusted performance, consistently achieving the highest values. Notably, the 3-agent variant exhibits unstable behavior with a sharp collapse near year-end, confirming that additional agents introduce optimization complexity without proportional benefits. The static 0.5/0.5 baseline closely tracks homogeneous methods, indicating that MARS-DA’s gains stem from learned dynamic weighting rather than simple ensembling.

\section{Conclusion}
\label{sec:conclusion}
This paper presented MARS-DA, a hierarchical reinforcement learning framework that decouples bidding strategy learning from market regime adaptation. By orchestrating specialized base agents through a dynamic Meta-Controller, MARS-DA overcomes the brittleness of monolithic RL policies and rigid optimization models. Extensive evaluations on PJM market data demonstrate that MARS-DA achieves superior risk-adjusted returns.

Crucially, the framework exhibited robust generalization during out-of-distribution stress tests, where it was the only agent to avoid catastrophic losses. The analysis revealed that this resilience stems from the Meta-Controller's learned ``contrarian" hedging strategy, which automatically reduces speculative exposure during extreme volatility. These results establish MARS-DA as a viable path toward autonomous, safe, and profitable participation in increasingly volatile renewable-dominated power markets.


\bibliographystyle{named}
\bibliography{ijcai26}

\end{document}